\documentclass[showpacs,amsmath,superscriptaddress,
reprint,aps]{revtex4-1}

\usepackage{hyperref}
\usepackage{graphicx}

\begin{document}
\title
{High efficiency dark-to-bright exciton conversion in carbon nanotubes}
\author{A.~Ishii}
\affiliation{Nanoscale Quantum Photonics Laboratory, RIKEN Cluster for Pioneering Research, Saitama 351-0198, Japan}
\affiliation{Quantum Optoelectronics Research Team, RIKEN Center for Advanced Photonics, Saitama 351-0198, Japan}
\author{H.~Machiya}
\affiliation{Nanoscale Quantum Photonics Laboratory, RIKEN Cluster for Pioneering Research, Saitama 351-0198, Japan}
\affiliation{Department of Electrical Engineering, The University of Tokyo, Tokyo 113-8656, Japan}
\author{Y.~K.~Kato}
\email[Corresponding author. ]{yuichiro.kato@riken.jp}
\affiliation{Nanoscale Quantum Photonics Laboratory, RIKEN Cluster for Pioneering Research, Saitama 351-0198, Japan}
\affiliation{Quantum Optoelectronics Research Team, RIKEN Center for Advanced Photonics, Saitama 351-0198, Japan}

\begin{abstract}
We report that dark excitons can have a large contribution to the emission intensity in carbon nanotubes due to an efficient exciton conversion from a dark state to a bright state. Time-resolved photoluminescence measurements are used to investigate decay dynamics and diffusion properties of excitons, and we obtain intrinsic lifetimes and diffusion lengths of bright excitons as well as diffusion coefficients for both bright and dark excitons. We find that the dark-to-bright transition rates can be considerably high, and that more than half of the dark excitons can be converted into the bright excitons. The state transition rates have a large chirality dependence with a family pattern, and the conversion efficiency is found to be significantly enhanced by adsorbed air molecules on the surface of the nanotubes. Our findings show the nontrivial significance of the dark excitons on the emission kinetics in low dimensional materials, and demonstrate the potential for engineering the dark-to-bright conversion process by using surface interactions.
\end{abstract}

\maketitle

\section{Introduction}
Due to limited screening of the Coulomb interaction, electron-hole pairs form tightly-bound excitons in atomically thin semiconductors such as transition metal dichalcogenides and carbon nanotubes (CNTs) \cite{Amori:2018, Mueller:2018}. The stable excitonic states play a central role in the optical processes, leading to unique phenomena such as quantum light emission \cite{He:2017, Ishii:2017, Tonndorf:2015}, exciton Hall effect \cite{Onga:2017}, and formation of interlayer excitons \cite{Rivera:2015}. In addition, there exist excitonic fine structures within the large binding energy, many of which are dark states with optical transitions forbidden by spin, valley, and parity selection rules \cite{Perebeinos:2005, Malic:2018}. The dark states, although they may seem subtle because of the lack of any spectral signatures, can lead to unexpected optical phenomena and are important for understanding of the exciton physics \cite{Stich:2014, Nishihara:2015, Zhang:2017, Zhou:2017}.

In particular, the parity-even dark states of excitons in CNTs are known to be responsible for the diminishing emission quantum efficiencies at cryogenic temperatures \cite{Mortimer:2007, Matsunaga:2009}, and various approaches have been taken to circumvent the ``dark'' nature of these states. For example, same-parity transitions by two-photon excitation \cite{Wang:2005, Maultzsch:2005} and intraexcitonic excitation by terahertz spectroscopy \cite{Luo:2015, Luo:2019} can address the dark states without relying on dipole transitions to and from the ground state, whereas magnetic and electric fields can be used to mix the bright and dark states \cite{Matsunaga:2008, Srivastava:2008, Uda:2016}. In the absence of external fields, light emission should not be affected by the presence of dark excitons, and indeed, time-resolved photoluminescence (PL) measurements show that the contribution of the dark excitons to PL intensity is insignificant \cite{Berciaud:2008, Gokus:2010, Hertel:2010, Crochet:2012}.

Here we show that conversion efficiency of the dark excitons to the bright excitons can become higher than 50\%, and that a considerable fraction of light emission can originate from the dark states. The kinetics of the bright and dark excitons in individual air-suspended CNTs are investigated using time-resolved PL measurements, and we determine the effective decay lifetimes and diffusion properties by measuring CNTs with various suspended lengths. We find that the dark excitons have intrinsic lifetimes exceeding our measurement capability, and their decay rates are solely limited by quenching at the tube ends. For long tubes, the dark exciton effective lifetime becomes comparable to the bright-dark transition time, and conversion to the bright excitons occurs efficiently. The transition rates exhibit a large chirality dependence with a family pattern, and we find that it can be significantly enhanced by adsorbed molecules on the CNTs.

\section{Photoluminescence spectroscopy and time-resolved measurements}

Our air-suspended carbon nanotubes are grown over trenches on Si substrates with metal pads, where molecular desorbed states of carbon nanotubes \cite{Uda:2018a} can be observed for an extended period of time \cite{SupplementalMaterials}. We perform electron-beam lithography and dry etching to form trenches with widths ranging from 0.5 to 4.0~$\mu$m. The samples are then oxidized in an annealing furnace at 1050$^\circ$C for an hour in order to form a 65-nm-thick oxide layer. Another lithography step defines the metal pad areas 50 $\mu$m away from the trenches, and sputtering is used to deposit 1.5~nm Ti and 40~nm Pt. Catalyst areas are patterned on the oxide layer near the trenches with a third lithography step, and 0.2-nm thermally evaporated Fe film is deposited and lifted off. Single-walled carbon nanotubes are synthesized by alcohol chemical vapor deposition at 800$^\circ$C with a growth time of 1~min \cite{Ishii:2015}.

PL measurements are performed with a homebuilt sample-scanning confocal microscopy system \cite{Ishii:2015, Ishii:2018}. We use a wavelength-tunable Ti:sapphire laser where the output can be switched between continuous-wave (CW) and $\sim$100-fs pulses with a repetition rate of 76~MHz. An excitation laser beam with power $P$ and wavelength $\lambda_\mathrm{ex}$ is focused onto the sample by an objective lens with a numerical aperture of 0.85 and a focal length of 1.8~mm, resulting in a spot size of $\sim$1~$\mu$m. The wavelength-dependent diameter of the focused laser has been characterized by performing PL line scans perpendicular to a suspended CNT. PL and the reflected beam are collected by the same objective lens and separated by a dichroic filter. A Si photodiode detects the reflected beam for imaging, while a translating mirror is used to switch between PL spectroscopy and time-resolved PL measurements. PL spectra are measured with an InGaAs photodiode array attached to a spectrometer, and time-resolved PL measurements are performed using a fiber-coupled superconducting single photon detector and a time-correlated single-photon counting module. The detection wavelength dependent instrument response function (IRF) is obtained by using super-continuum white light pulses dispersed by a spectrometer. All measurements are conducted at room temperature in a nitrogen-purged environment.

\begin{figure}
\includegraphics{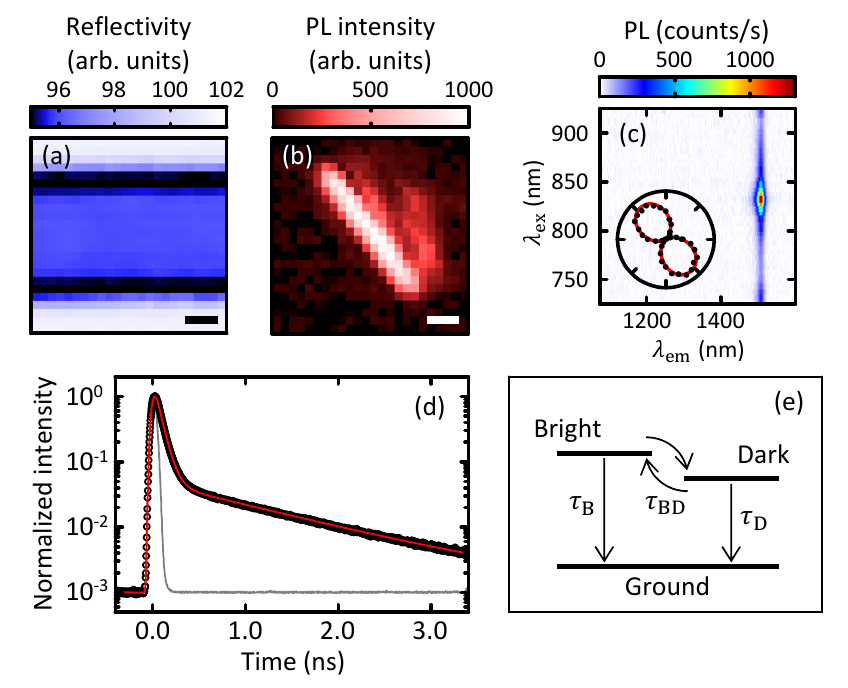}
\caption{
\label{Fig1} (a) and (b) Reflectivity and PL images, respectively. $\lambda_\mathrm{ex}=830$~nm and $P=2$~$\mu$W are used, and the PL image is extracted within a window of 10~nm centered at 1469~nm with integration window. The weaker emission to the right side of the tube is a ghost image arising from excitation laser reflecting at the bottom of the trench. The scale bars are 1~$\mu$m. (c) A PLE map taken with $P=2$~$\mu$W. The inset shows polarization dependence of PL intensity. The red line is a sine fit. (d) PL decay curve measured under pulsed excitation with $\lambda_\mathrm{ex}=825$~nm and $P=5$~nW. The gray line represents the IRF, and the red curve is the bi-exponential fit. (e) Schematic of the three-level model for exciton decay dynamics.}
\end{figure}

For characterizing as-grown suspended CNTs, we perform PL spectroscopy measurements under CW excitation \cite{Ishii:2015}. Line scans along the trenches are used to locate the suspended CNTs, and we take reflectivity and PL images to confirm that they are fully suspended. Figures~\ref{Fig1}(a) and (b) show reflectivity and PL images, respectively, for a relatively long tube. The PL image at the emission wavelength $\lambda_\mathrm{em}$ shows a smooth spatial profile, indicating that the suspended nanotube is defect free and does not contain any quenching sites or trapping sites. Next, we perform PL excitation (PLE) spectroscopy [Fig.~\ref{Fig1}(c)] to identify the chirality $(n,m)=(11,7)$ by comparing the $E_{11}$ and $E_{22}$ wavelengths to tabulated data \cite{Ishii:2015}. Polarization dependence of PL intensity is then measured to determine the angle of the nanotube [inset of Fig.~\ref{Fig1}(c)], and the suspended length $L=4.8$~$\mu$m is calculated using the angle and the designed trench width.

Using such well-characterized CNTs, we perform time-resolved PL measurements \cite{Ishii:2018} to investigate exciton decay dynamics by mode locking the laser and collecting PL from the center of the nanotube. The excitation laser is tuned to the $E_{22}$ wavelength and the polarization is aligned parallel to the CNT axis unless otherwise noted. Figure~\ref{Fig1}(d) shows a PL decay curve taken from the same CNT measured in Figs.~\ref{Fig1}(a-c). The decay curve shows two components with different lifetimes, and a fit is done with a bi-exponential function convoluted with the IRF. The bi-exponential function is defined as $A_1 \exp(-t / \tau_1) + A_2 \exp(-t / \tau_2)$, where $t$ is time and $\tau_1<\tau_2$, and we obtain the decay lifetimes $\tau_1=67$~ps and $\tau_2=1216$~ps as well as the intensity fraction of the slow component $Y_2 = A_2 \tau_2 / (A_1 \tau_1 + A_2 \tau_2) = 36.6\%$ from the fit. It is surprising that the slow decay component constitutes such a large fraction of the PL intensity, because it is believed to originate from dark excitons converted into bright excitons \cite{Berciaud:2008, Gokus:2010, Crochet:2012}.

The conversion kinetics can be understood by analyzing the bi-exponential decay curve with a three-level model \cite{Gokus:2010}, which is schematically shown in Fig.~\ref{Fig1}(e). The rate equations for the populations of bright excitons $N_\mathrm{B}$ and dark excitons $N_\mathrm{D}$ are expressed as
\begin{eqnarray}
\frac{dN_\mathrm{B}}{dt} = -\bigl( \frac{1}{\tau_\mathrm{B}}+\frac{1}{\tau_\mathrm{BD}}\bigr) N_\mathrm{B} + \frac{1}{\tau_\mathrm{BD}} N_\mathrm{D},\\
\frac{dN_\mathrm{D}}{dt} = -\bigl( \frac{1}{\tau_\mathrm{D}}+\frac{1}{\tau_\mathrm{BD}}\bigr) N_\mathrm{D} + \frac{1}{\tau_\mathrm{BD}} N_\mathrm{B},
\end{eqnarray}
where $\tau_\mathrm{B}$ and $\tau_\mathrm{D}$ are the effective decay lifetimes for bright and dark states, respectively, and $\tau_\mathrm{BD}$ is the state-transition time between the bright and the dark states. By solving these kinetic equations assuming an initial population ratio of $N_\mathrm{B}:N_\mathrm{D}=1:1$, a bi-exponential function representing the time-evolution of $N_\mathrm{B}$ can be obtained, which corresponds to the observed PL decay curves. From the experimentally obtained parameters $\tau_1, \tau_2, A_1$, and $A_2$, we obtain $\tau_\mathrm{B}=69$~ps and $\tau_\mathrm{D}=2422$~ps as well as $\tau_\mathrm{BD}=2369$~ps. The conversion efficiency $\eta_\uparrow$ is given by $\tau_\mathrm{D}/(\tau_\mathrm{D}+\tau_\mathrm{BD})=51$\%, showing that more than half of the dark excitons are converted to the bright excitons.

\begin{figure*}
\includegraphics{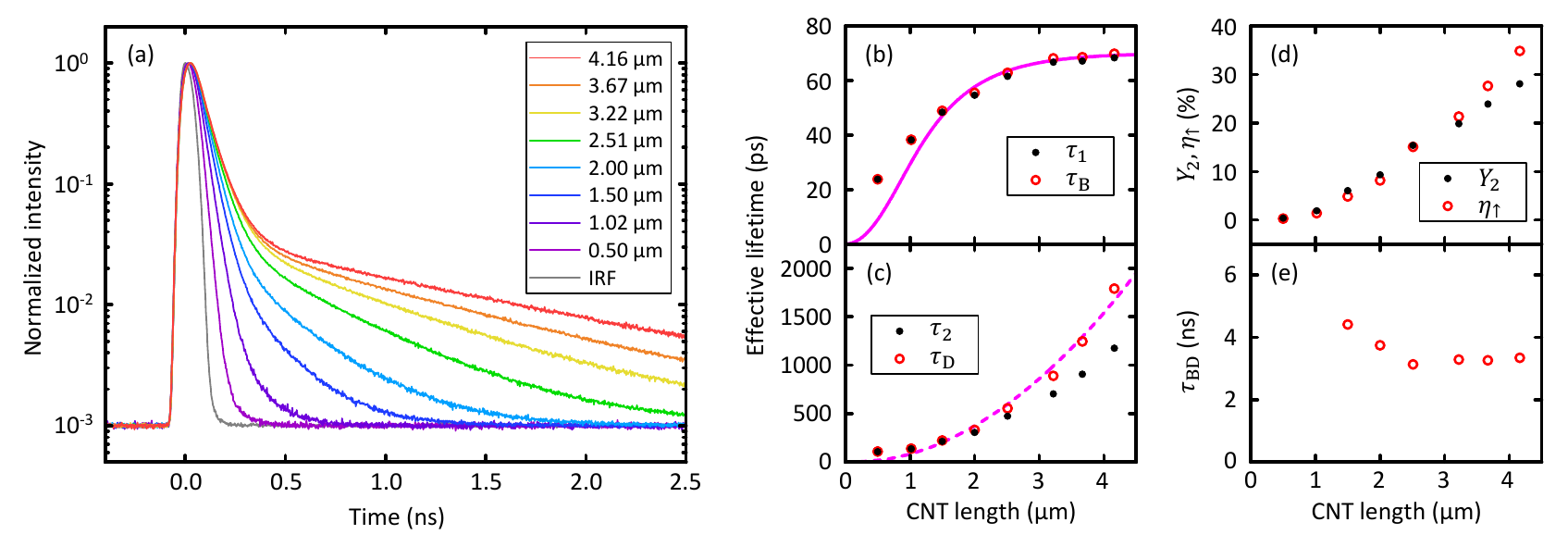}
\caption{
\label{Fig2} (a) PL decay curved measured from (9,8) nanotubes with various lengths ranging from 0.5 to 4.2~$\mu$m. The gray line is the IRF. $P=5$~nW and laser polarization parallel to the CNT axis are used. (b) Length dependence of $\tau_1$ (filled circles) and $\tau_\mathrm{B}$(open circles). The solid line is the fit with Eq.~\ref{tauBright}. (c) Length dependence of $\tau_2$ (filled circles) and $\tau_\mathrm{D}$ (open circles). The broken line is the fit with Eq.~\ref{tauDark}. (d) Length dependence of $Y_2$ (filled circles) and $\eta_\uparrow$ (open circles). (e) Length dependence of $\tau_\mathrm{BD}$. In (b) and (c), data points for $L \le 1.3$~$\mu$m are excluded from the fits.}
\end{figure*}

\section{Diffusion-limited conversion efficiency}

It turns out that the dark-to-bright conversion efficiency depends strongly on the suspended length. We select (9,8) CNTs with lengths ranging from 0.5 to 4.2~$\mu$m and measure PL decay curves with $E_{22}$ excitation at a low power so that the effects of exciton-exciton annihilation can be negligible \cite{Ishii:2015}. The influence of excitation and detection wavelengths as well as the excitation power dependence on the decay dynamics are discussed in the Supplemental Materials \cite{SupplementalMaterials}. In Fig.~\ref{Fig2}(a), the PL decay curves taken at various suspended lengths are shown, where drastic changes are observed. As CNT length becomes longer, decay lifetimes of both fast and slow components become longer, and the intensity fraction of the slow component increases. The decay curves are fitted with the bi-exponential function, and the length dependence of $\tau_1$, $\tau_2$, and $Y_2$ are plotted as filled circles in Figs.~\ref{Fig2}(b-d), respectively. For each CNT, we calculate $\tau_\mathrm{B}$, $\tau_\mathrm{D}$, $\eta_\uparrow$, and $\tau_\mathrm{BD}$ as shown by open circles in Figs.~\ref{Fig2}(b-e), respectively, based on the three-level model. The values of $\tau_\mathrm{B}$ are very close to $\tau_1$, while $\tau_\mathrm{D}$ considerably deviates from $\tau_2$ for long CNTs, indicating that $\tau_\mathrm{BD}$ is important for dark exciton dynamics. Considering the uncertainty of the measurement results for short CNTs where the decay is very fast, $\tau_\mathrm{BD}$ for longer tubes should be considered as the intrinsic values [Fig.~\ref{Fig2}(e)]. In contrast to $\tau_\mathrm{B}$ which is almost saturated at CNT lengths longer than 3~$\mu$m, $\tau_\mathrm{D}$ continues to grow even at the longest length, resulting in the steady increase of $Y_2$ and $\eta_\uparrow$ with the CNT length.

We analyze the length dependence of the effective lifetimes with a random walk theory where the influence of end quenching is evaluated \cite{Ishii:2015, Anderson:2013}. As the end quenching efficiency depends on the suspended length $L$ and the generated position $z$ of excitons, the effective lifetime is given by
\begin{equation}
\label{tauEff}
\tau(L) = \frac{\int_{-L/2}^{L/2} g(z) \tau^*(L,z) dz}{\int_{-L/2}^{L/2} g(z) dz},
\end{equation}
where $g(z)$ is the normalized Gaussian profile representing the excitation laser profile with a wavelength-dependent spot size. The behavior of $\tau(L)$ is mostly determined by the position-specific lifetime $\tau^*(L,z) = \int_{0}^{\infty} S_\mathrm{I}(t) S_\mathrm{E}(L,D,z,t) dt$, where $S_\mathrm{I}(t)=\exp(-t/\tau^\mathrm{int})$ is the survival probability through intrinsic decay with a lifetime of $\tau^\mathrm{int}$, and $S_\mathrm{E}(L,D,z,t)$ is the end quenching survival probability for excitons generated at $z$ with diffusion coefficient $D$.

For a finite $\tau^\mathrm{int}$, we can simplify the expression to \cite{Ishii:2015}
\begin{equation}
\label{tauBright}
\tau^*(L,z) = \tau^\mathrm{int} \biggl\{1 - \frac{\cosh(z / l)}{\cosh (L / 2l)}\biggr\},
\end{equation}
where $l_\mathrm{B}=\sqrt{D \tau^\mathrm{int}}$ is the exciton diffusion length, and this model is adopted for the length dependence of $\tau_\mathrm{B}$. We obtain the intrinsic lifetime $\tau_\mathrm{B}^\mathrm{int}=70$~ps and diffusion length $l_\mathrm{B}=390$~nm as well as $D=22$~cm$^2$/s from the fit [solid line in Fig.~\ref{Fig2}(b)]. The intrinsic lifetime is almost the same as the reported values \cite{Xiao:2010, Crochet:2012}, while the diffusion length is shorter than those reported for similar air-suspended CNTs \cite{Ishii:2015, Moritsubo:2010}. We note that this analysis gives the true diffusion length for the bright excitons, while the previously reported values include the dark exciton effects.

For the dark excitons, in contrast, fits by Eq.~\ref{tauBright} does not converge because $\tau_\mathrm{D}$ does not show any sign of saturation, implying that the intrinsic lifetime for dark excitons is much longer than the effective lifetime limited by end quenching. Letting $\tau^\mathrm{int} \rightarrow \infty$,
\begin{equation}
\label{tauDark}
\tau^*(L,z) = \frac{L^2-4z^2}{8D},
\end{equation}
explaining the rapid increase of $\tau_\mathrm{D}$ with $L$ responsible for the enhanced conversion efficiency. The diffusion coefficient $D$ can be determined by fitting the length dependence [broken line in Fig.~\ref{Fig2}(c)], and we obtain $D=13$~cm$^2$/s.

\section{Chirality dependence of exciton dynamics and the state transition rate}

\begin{figure}
\includegraphics{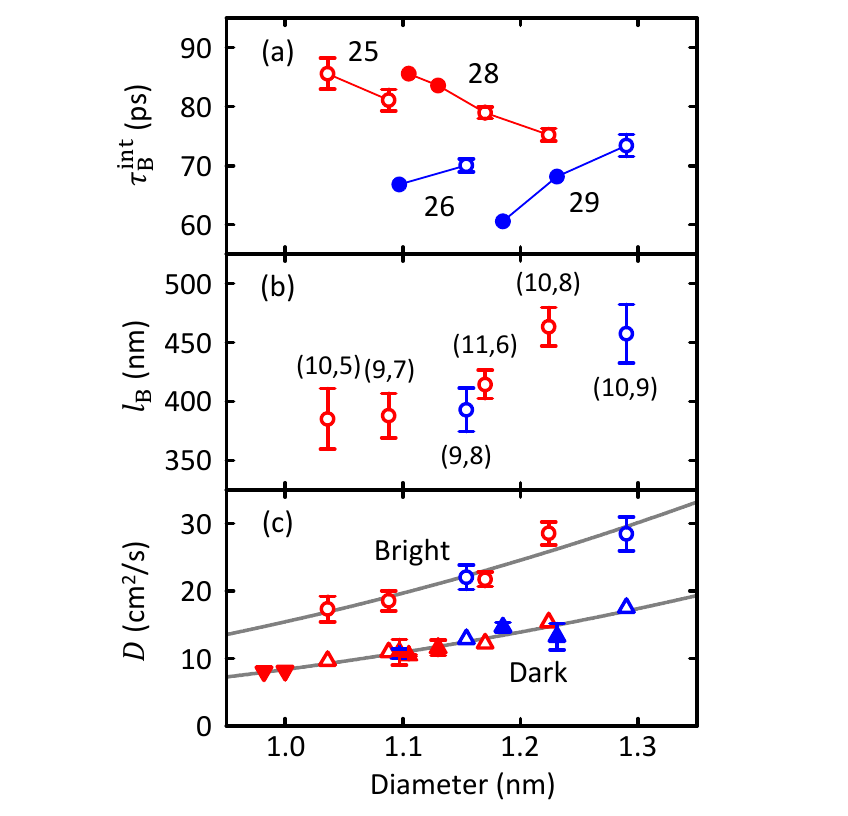}
\caption{
\label{Fig3} (a) Diameter dependence of $\tau_\mathrm{B}^\mathrm{int}$. some more details on data criteria. Open circles are obtained from the length-dependence fit, and filled circles are the longest effective lifetimes measured within the same chiralities. Chiralities with the same family number ($2n+m$) are connected, and the family numbers are displayed. (b) Diameter dependence of $l_\mathrm{B}$. The chiral indices are shown for each data point. (c) Diameter dependence of $D$ for bright excitons (open circles) and dark excitons (triangles). The filled triangles are estimated from a limited number of CNTs, where up and down triangles are obtained from two and one data points, respectively. Gray lines are fit results with a power function. (a-c) Red and blue symbols correspond to type 1 and 2 of chiralities, respectively, and the error bars show the standard error of the fits. Error bars for the open triangles in (c) are not shown as they are smaller than the symbols.}
\end{figure}

We now investigate the chirality dependence of the dark-to-bright conversion efficiency. The same measurements for different chiralities are repeated, and we first characterize the exciton diffusion properties. Figures~\ref{Fig3}(a) and (b) show $\tau_\mathrm{B}^\mathrm{int}$ and $l_\mathrm{B}$, respectively, as a function of CNT diameter $d$. Since $\tau_\mathrm{B}$ is mostly saturated at long CNT lengths [Fig. 2(b)], we also plot the largest values of $\tau_\mathrm{B}$ for 5 additional chiralities for which multiple CNTs longer than 3~$\mu$m are measured [filled circles in Fig.~\ref{Fig3}(a)]. We find a clear family pattern dispersion for $\tau_\mathrm{B}^\mathrm{int}$ ranging from 61~ps to 86~ps, whereas $l_\mathrm{B}$ shows an increasing trend with the CNT diameter.

The diffusion constant $D$ for the bright excitons is obtained from $\tau_\mathrm{B}^\mathrm{int}$ and $l_\mathrm{B}$ [open circles in Fig.~\ref{Fig3}(c)], while $D$ of the dark excitons is directly obtained from the length dependence fits [open triangles in Fig.~\ref{Fig3}(c)]. In all the chiralities, Eq.~\ref{tauDark} accurately reproduces the length dependence of $\tau_\mathrm{D}$, indicating that exciton diffusion limits the effective lifetime for the dark excitons. In addition, the fits for the length dependent $\tau_\mathrm{D}$ can converge with a limited number of data, allowing us to plot $D$ of the dark excitons for 8 additional chiralities [filled triangles in Fig.~\ref{Fig3}(c)].

The diffusion coefficients for both the bright and dark states show clear diameter $d$ dependence, and we use a power law function $D=D_0 (d/d_0)^\alpha$ to fit the data, where $D_0$ is the diffusion coefficient at the diameter $d_0=1$~nm and $\alpha$ is the exponent. The fit results show a good agreement with the experimental data [gray lines in Fig.~\ref{Fig3}(c)], where we obtain $D_0=15.36\pm1.24$~cm$^2$/s and $\alpha=2.56\pm0.43$ for bright excitons and $D_0=8.36\pm0.31$~cm$^2$/s and $\alpha=2.79\pm0.22$ for dark excitons. The results are consistent with the theoretical prediction of $D \propto d^{2.5}$, where the chirality dependence of the effective mass of excitons and the exciton-phonon scattering rate is considered \cite{Srivastava:2009}. We find that $D_0$ differs by about a factor of 2 between bright and dark excitons, which is reasonable because the effective mass of the bright excitons is much smaller than that of dark excitons due to the difference of their band structures \cite{Perebeinos:2005}. A quantitative comparison of the theory with the experimental data, however, is still challenging due to other mechanisms of exciton scattering \cite{Crochet:2012}.

\begin{figure}
\includegraphics{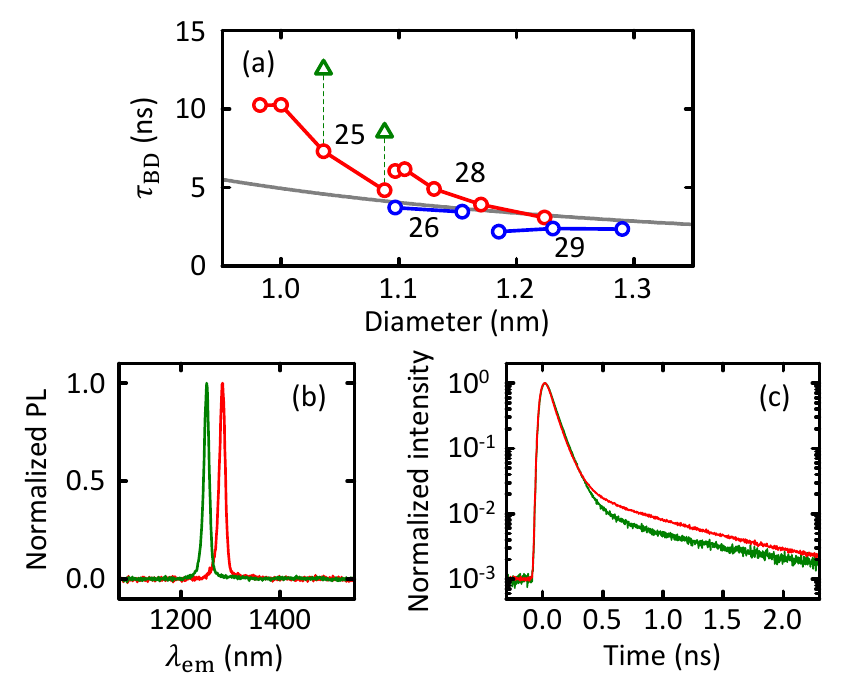}
\caption{
\label{Fig4} (a) Diameter dependence of $\tau_\mathrm{BD}$. The data are taken from tubes with $L \ge 2.5$~$\mu$m and with $P \le 2$~nW. Chiralities with a same family number are connected, and the family numbers are displayed. Red and blue circles correspond to chiralities which belong to type 1 and type 2, respectively. Gray line is a fit described in the text. Green triangles are obtained from molecular desorbed state of (9,7) and (10,5) CNTs. (b) and (c) Normalized PL spectra and PL decay curves, respectively, of a same (9,7) CNT. Red and green lines correspond to molecular adsorbed and desorbed states, respectively.}
\end{figure}

Finally we discuss the transition rate between the bright and dark states. As we have observed in Fig.~\ref{Fig2}(e), the transition time $\tau_\mathrm{BD}$ does not depend on CNT length and is more reliable for long CNTs, and therefore we can obtain the intrinsic values from a relatively long CNT for each chirality. In Fig.~\ref{Fig4}(a), such representative values of $\tau_\mathrm{BD}$ are plotted as a function of CNT diameter, and a clear family pattern on top of a decreasing trend with CNT diameter is observed. It is worth mentioning that such a wide variation of $\tau_\mathrm{BD}$ considerably affects the observed PL decay curves, particularly in the intensity ratio between fast and slow decay components. For example, 2.5-$\mu$m-long (12,1) nanotube with $\tau_\mathrm{BD}=10.24$~ns shows $Y_2=8.6$\% and $\eta_\uparrow=8.5$\%, while a (10,9) CNT with the same length has $\tau_\mathrm{BD}=2.20$~ns exhibiting $Y_2=18.2$\% and $\eta_\uparrow=17.4$\%. The chirality dependence of the bright-dark transition rate is likely related to the bright-dark energy separation $\Delta E$, where a similar family pattern is known with a leading order dependence of $\Delta E \propto 1/d^2$ \cite{Capaz:2006}. To confirm this dependence, we take near-armchair chiralities from each family group, and a fit is done by assuming $\Delta E=3$~meV at $d=1$~nm \cite{Matsunaga:2008}. The fit result is displayed as a gray line in Fig.~\ref{Fig4}(a), showing a fairly good agreement with the data.

As the state transition requires a parity flip process, it should be possible to control the conversion efficiency by modifying the exciton scattering site density. We compare the state transition time for a nanotube with and without molecular adsorption \cite{SupplementalMaterials}. As shown in Fig.~\ref{Fig4}(b), a blueshift by 32~nm is observed in the PL spectrum of a (9,7) CNT, due to the change of the molecular screening effect \cite{Uda:2018a}. The PL decay curves are taken before and after molecular desorption as shown in Fig.~\ref{Fig4}(c), and we observe a drastic change of the state transition rate where $\tau_\mathrm{BD}$ increases by a factor of 2, resulting in a decrease of $Y_2$ and $\eta_\uparrow$ from 16.8\% to 9.1\% and from 17.3\% to 8.7\%, respectively. We plot $\tau_\mathrm{BD}$ of the desorbed state for (9,7) and (10,5) CNTs as green triangles in Fig.~\ref{Fig4}(a). We note that $\tau_\mathrm{BD}$ can be affected by the change of $\Delta E$ due to the reduced molecular screening, but it only accounts for 10\% of the $\tau_\mathrm{BD}$ reduction if we assume that $\Delta E$ scales similar to the $E_{11}$ energy \cite{Lefebvre:2008}. In addition, we observe 8.9\% decrease of $\tau_\mathrm{D}$ after the molecular desorption. If $\tau_\mathrm{D}$ is still determined by end quenching as described by Eq.~\ref{tauDark}, the reduction of $\tau_\mathrm{D}$ implies 9.7\% increase of diffusion coefficient, consistent with the reduced scattering.

\section{Conclusion}
We have investigated exciton dynamics in air-suspended CNTs and have shown that the dark excitons can be converted to the bright state with efficiencies higher than 50\%. Time-resolved PL measurements reveal the decay dynamics of excitons, where the bright excitons have intrinsic lifetimes ranging from 61 to 86 ps, while intrinsic lifetimes of the dark excitons are found to be too large to experimentally determine. For long CNTs, the dark exciton decay rate becomes comparable to the bright-dark state transition rate, resulting in a considerable contribution of the dark excitons to the emission intensity. Reflecting the dissimilar lifetimes, the diffusion properties are contrastingly different for bright and dark excitons. By understanding the diffusive motion of the dark states, it should become possible to improve the performance of nanotube single photon emitters as they utilize the mobile nature of excitons \cite{Ma:2015prl, Ma:2015NatNano, He:2017, Ishii:2017}. The distinct difference between bright and dark excitons are also expected in other atomically thin materials, which may potentially be used to optimize exciton transport in van der Waals heterostructure photovoltaic devices \cite{Furchi:2014, Memaran:2015}. We also show chirality dependence of the bright-dark state transition rate, where a wide range of variation with a clear family pattern is observed. Furthermore, by comparing exciton dynamics for pristine and molecular adsorbed states of the same CNTs, we find that the state transition rate is significantly reduced by molecular desorption, suggesting an enhanced parity flip process originating from the adsorbed molecules. If manipulation of multiple exciton species becomes possible by engineering the exciton conversion process, it may lead to development of advanced photonic and optoelectronic devices with devoted channels for exciton transport, recombination, and dissociation.

\begin{acknowledgments}
Work supported by JSPS (KAKENHI JP16K13613 and JP17H07359), MEXT (Nanotechnology Platform), and RIKEN (Incentive Research Project). H.M. acknowledges support by JSPS (Research Fellowship for Young Scientists) and RIKEN (Junior Research Associate Program). We acknowledge the Advanced Manufacturing Support Team at RIKEN for technical assistance.
\end{acknowledgments}


\begin{thebibliography}{40}%
\makeatletter
\providecommand \@ifxundefined [1]{%
 \@ifx{#1\undefined}
}%
\providecommand \@ifnum [1]{%
 \ifnum #1\expandafter \@firstoftwo
 \else \expandafter \@secondoftwo
 \fi
}%
\providecommand \@ifx [1]{%
 \ifx #1\expandafter \@firstoftwo
 \else \expandafter \@secondoftwo
 \fi
}%
\providecommand \natexlab [1]{#1}%
\providecommand \enquote  [1]{``#1''}%
\providecommand \bibnamefont  [1]{#1}%
\providecommand \bibfnamefont [1]{#1}%
\providecommand \citenamefont [1]{#1}%
\providecommand \href@noop [0]{\@secondoftwo}%
\providecommand \href [0]{\begingroup \@sanitize@url \@href}%
\providecommand \@href[1]{\@@startlink{#1}\@@href}%
\providecommand \@@href[1]{\endgroup#1\@@endlink}%
\providecommand \@sanitize@url [0]{\catcode `\\12\catcode `\$12\catcode
  `\&12\catcode `\#12\catcode `\^12\catcode `\_12\catcode `\%12\relax}%
\providecommand \@@startlink[1]{}%
\providecommand \@@endlink[0]{}%
\providecommand \url  [0]{\begingroup\@sanitize@url \@url }%
\providecommand \@url [1]{\endgroup\@href {#1}{\urlprefix }}%
\providecommand \urlprefix  [0]{URL }%
\providecommand \Eprint [0]{\href }%
\providecommand \doibase [0]{https://doi.org/}%
\providecommand \selectlanguage [0]{\@gobble}%
\providecommand \bibinfo  [0]{\@secondoftwo}%
\providecommand \bibfield  [0]{\@secondoftwo}%
\providecommand \translation [1]{[#1]}%
\providecommand \BibitemOpen [0]{}%
\providecommand \bibitemStop [0]{}%
\providecommand \bibitemNoStop [0]{.\EOS\space}%
\providecommand \EOS [0]{\spacefactor3000\relax}%
\providecommand \BibitemShut  [1]{\csname bibitem#1\endcsname}%
\let\auto@bib@innerbib\@empty
\bibitem [{\citenamefont {Amori}\ \emph {et~al.}(2018)\citenamefont {Amori},
  \citenamefont {Hou},\ and\ \citenamefont {Krauss}}]{Amori:2018}%
  \BibitemOpen
  \bibfield  {author} {\bibinfo {author} {\bibfnamefont {A.~R.}\ \bibnamefont
  {Amori}}, \bibinfo {author} {\bibfnamefont {Z.}~\bibnamefont {Hou}}, \ and\
  \bibinfo {author} {\bibfnamefont {T.~D.}\ \bibnamefont {Krauss}},\ }\bibfield
   {title} {\bibinfo {title} {Excitons in single-walled carbon nanotubes and
  their dynamics},\ }\href {\doibase 10.1146/annurev-physchem-050317-014241}
  {\bibfield  {journal} {\bibinfo  {journal} {Annu. Rev. Phys. Chem.}\ }\textbf
  {\bibinfo {volume} {69}},\ \bibinfo {pages} {81} (\bibinfo {year}
  {2018})}\BibitemShut {NoStop}%
\bibitem [{\citenamefont {Mueller}\ and\ \citenamefont
  {Malic}(2018)}]{Mueller:2018}%
  \BibitemOpen
  \bibfield  {author} {\bibinfo {author} {\bibfnamefont {T.}~\bibnamefont
  {Mueller}}\ and\ \bibinfo {author} {\bibfnamefont {E.}~\bibnamefont
  {Malic}},\ }\bibfield  {title} {\bibinfo {title} {Exciton physics and device
  application of two-dimensional transition metal dichalcogenide
  semiconductors},\ }\href {\doibase 10.1038/s41699-018-0074-2} {\bibfield
  {journal} {\bibinfo  {journal} {npj 2D Mater. Appl.}\ }\textbf
  {\bibinfo {volume} {2}},\ \bibinfo {pages} {29} (\bibinfo {year}
  {2018})}\BibitemShut {NoStop}%
\bibitem [{\citenamefont {He}\ \emph {et~al.}(2017)\citenamefont {He},
  \citenamefont {Hartmann}, \citenamefont {Ma}, \citenamefont {Kim},
  \citenamefont {Ihly}, \citenamefont {Blackburn}, \citenamefont {Gao},
  \citenamefont {Kono}, \citenamefont {Yomogida}, \citenamefont {Hirano},
  \citenamefont {Tanaka}, \citenamefont {Kataura}, \citenamefont {Htoon},\ and\
  \citenamefont {Doorn}}]{He:2017}%
  \BibitemOpen
  \bibfield  {author} {\bibinfo {author} {\bibfnamefont {X.}~\bibnamefont
  {He}}, \bibinfo {author} {\bibfnamefont {N.~F.}\ \bibnamefont {Hartmann}},
  \bibinfo {author} {\bibfnamefont {X.}~\bibnamefont {Ma}}, \bibinfo {author}
  {\bibfnamefont {Y.}~\bibnamefont {Kim}}, \bibinfo {author} {\bibfnamefont
  {R.}~\bibnamefont {Ihly}}, \bibinfo {author} {\bibfnamefont {J.~L.}\
  \bibnamefont {Blackburn}}, \bibinfo {author} {\bibfnamefont {W.}~\bibnamefont
  {Gao}}, \bibinfo {author} {\bibfnamefont {J.}~\bibnamefont {Kono}}, \bibinfo
  {author} {\bibfnamefont {Y.}~\bibnamefont {Yomogida}}, \bibinfo {author}
  {\bibfnamefont {A.}~\bibnamefont {Hirano}}, \bibinfo {author} {\bibfnamefont
  {T.}~\bibnamefont {Tanaka}}, \bibinfo {author} {\bibfnamefont
  {H.}~\bibnamefont {Kataura}}, \bibinfo {author} {\bibfnamefont
  {H.}~\bibnamefont {Htoon}}, \ and\ \bibinfo {author} {\bibfnamefont {S.~K.}\
  \bibnamefont {Doorn}},\ }\bibfield  {title} {\bibinfo {title} {Tunable
  room-temperature single-photon emission at telecom wavelengths from $sp^3$
  defects in carbon nanotubes},\ }\href
  {http://dx.doi.org/10.1038/nphoton.2017.119} {\bibfield  {journal} {\bibinfo
  {journal} {Nat. Photon.}\ }\textbf {\bibinfo {volume} {11}},\ \bibinfo
  {pages} {577} (\bibinfo {year} {2017})}\BibitemShut {NoStop}%
\bibitem [{\citenamefont {Ishii}\ \emph {et~al.}(2017)\citenamefont {Ishii},
  \citenamefont {Uda},\ and\ \citenamefont {Kato}}]{Ishii:2017}%
  \BibitemOpen
  \bibfield  {author} {\bibinfo {author} {\bibfnamefont {A.}~\bibnamefont
  {Ishii}}, \bibinfo {author} {\bibfnamefont {T.}~\bibnamefont {Uda}}, \ and\
  \bibinfo {author} {\bibfnamefont {Y.~K.}~\bibnamefont {Kato}},\ }\bibfield
  {title} {\bibinfo {title} {Room-temperature single-photon emission from
  micrometer-long air-suspended carbon nanotubes},\ }\href {\doibase
  10.1103/physrevapplied.8.054039} {\bibfield  {journal} {\bibinfo  {journal}
  {Phys. Rev. Applied}\ }\textbf {\bibinfo {volume} {8}},\ \bibinfo {pages}
  {054039} (\bibinfo {year} {2017})}\BibitemShut {NoStop}%
\bibitem [{\citenamefont {Tonndorf}\ \emph {et~al.}(2015)\citenamefont
  {Tonndorf}, \citenamefont {Schmidt}, \citenamefont {Schneider}, \citenamefont
  {Kern}, \citenamefont {Buscema}, \citenamefont {Steele}, \citenamefont
  {Castellanos-Gomez}, \citenamefont {van~der Zant}, \citenamefont
  {de~Vasconcellos},\ and\ \citenamefont {Bratschitsch}}]{Tonndorf:2015}%
  \BibitemOpen
  \bibfield  {author} {\bibinfo {author} {\bibfnamefont {P.}~\bibnamefont
  {Tonndorf}}, \bibinfo {author} {\bibfnamefont {R.}~\bibnamefont {Schmidt}},
  \bibinfo {author} {\bibfnamefont {R.}~\bibnamefont {Schneider}}, \bibinfo
  {author} {\bibfnamefont {J.}~\bibnamefont {Kern}}, \bibinfo {author}
  {\bibfnamefont {M.}~\bibnamefont {Buscema}}, \bibinfo {author} {\bibfnamefont
  {G.~A.}\ \bibnamefont {Steele}}, \bibinfo {author} {\bibfnamefont
  {A.}~\bibnamefont {Castellanos-Gomez}}, \bibinfo {author} {\bibfnamefont
  {H.~S.~J.}\ \bibnamefont {van~der Zant}}, \bibinfo {author} {\bibfnamefont
  {S.~M.}\ \bibnamefont {de~Vasconcellos}}, \ and\ \bibinfo {author}
  {\bibfnamefont {R.}~\bibnamefont {Bratschitsch}},\ }\bibfield  {title}
  {\bibinfo {title} {Single-photon emission from localized excitons in an
  atomically thin semiconductor},\ }\href {\doibase 10.1364/OPTICA.2.000347}
  {\bibfield  {journal} {\bibinfo  {journal} {Optica}\ }\textbf {\bibinfo
  {volume} {2}},\ \bibinfo {pages} {347} (\bibinfo {year} {2015})}\BibitemShut
  {NoStop}%
\bibitem [{\citenamefont {Onga}\ \emph {et~al.}(2017)\citenamefont {Onga},
  \citenamefont {Zhang}, \citenamefont {Ideue},\ and\ \citenamefont
  {Iwasa}}]{Onga:2017}%
  \BibitemOpen
  \bibfield  {author} {\bibinfo {author} {\bibfnamefont {M.}~\bibnamefont
  {Onga}}, \bibinfo {author} {\bibfnamefont {Y.}~\bibnamefont {Zhang}},
  \bibinfo {author} {\bibfnamefont {T.}~\bibnamefont {Ideue}}, \ and\ \bibinfo
  {author} {\bibfnamefont {Y.}~\bibnamefont {Iwasa}},\ }\bibfield  {title}
  {\bibinfo {title} {Exciton {H}all effect in monolayer {M}o{S}$_2$},\ }\href
  {https://doi.org/10.1038/nmat4996} {\bibfield  {journal} {\bibinfo  {journal}
  {Nat. Mater.}\ }\textbf {\bibinfo {volume} {16}},\ \bibinfo {pages} {1193}
  (\bibinfo {year} {2017})}\BibitemShut {NoStop}%
\bibitem [{\citenamefont {Rivera}\ \emph {et~al.}(2015)\citenamefont {Rivera},
  \citenamefont {Schaibley}, \citenamefont {Jones}, \citenamefont {Ross},
  \citenamefont {Wu}, \citenamefont {Aivazian}, \citenamefont {Klement},
  \citenamefont {Seyler}, \citenamefont {Clark}, \citenamefont {Ghimire},
  \citenamefont {Yan}, \citenamefont {Mandrus}, \citenamefont {Yao},\ and\
  \citenamefont {Xu}}]{Rivera:2015}%
  \BibitemOpen
  \bibfield  {author} {\bibinfo {author} {\bibfnamefont {P.}~\bibnamefont
  {Rivera}}, \bibinfo {author} {\bibfnamefont {J.~R.}\ \bibnamefont
  {Schaibley}}, \bibinfo {author} {\bibfnamefont {A.~M.}\ \bibnamefont
  {Jones}}, \bibinfo {author} {\bibfnamefont {J.~S.}\ \bibnamefont {Ross}},
  \bibinfo {author} {\bibfnamefont {S.}~\bibnamefont {Wu}}, \bibinfo {author}
  {\bibfnamefont {G.}~\bibnamefont {Aivazian}}, \bibinfo {author}
  {\bibfnamefont {P.}~\bibnamefont {Klement}}, \bibinfo {author} {\bibfnamefont
  {K.}~\bibnamefont {Seyler}}, \bibinfo {author} {\bibfnamefont
  {G.}~\bibnamefont {Clark}}, \bibinfo {author} {\bibfnamefont {N.~J.}\
  \bibnamefont {Ghimire}}, \bibinfo {author} {\bibfnamefont {J.}~\bibnamefont
  {Yan}}, \bibinfo {author} {\bibfnamefont {D.~G.}\ \bibnamefont {Mandrus}},
  \bibinfo {author} {\bibfnamefont {W.}~\bibnamefont {Yao}}, \ and\ \bibinfo
  {author} {\bibfnamefont {X.}~\bibnamefont {Xu}},\ }\bibfield  {title}
  {\bibinfo {title} {Observation of long-lived interlayer excitons in monolayer
  {M}o{S}e$_2$-{WS}e$_2$ heterostructures},\ }\href
  {https://doi.org/10.1038/ncomms7242} {\bibfield  {journal} {\bibinfo
  {journal} {Nat. Commun.}\ }\textbf {\bibinfo {volume} {6}},\ \bibinfo {pages}
  {6242} (\bibinfo {year} {2015})}\BibitemShut {NoStop}%
\bibitem [{\citenamefont {Perebeinos}\ \emph {et~al.}(2005)\citenamefont
  {Perebeinos}, \citenamefont {Tersoff},\ and\ \citenamefont
  {Avouris}}]{Perebeinos:2005}%
  \BibitemOpen
  \bibfield  {author} {\bibinfo {author} {\bibfnamefont {V.}~\bibnamefont
  {Perebeinos}}, \bibinfo {author} {\bibfnamefont {J.}~\bibnamefont {Tersoff}},
  \ and\ \bibinfo {author} {\bibfnamefont {P.}~\bibnamefont {Avouris}},\
  }\bibfield  {title} {\bibinfo {title} {Radiative lifetime of excitons in
  carbon nanotubes},\ }\href {\doibase 10.1021/nl051828s} {\bibfield  {journal}
  {\bibinfo  {journal} {Nano Lett.}\ }\textbf {\bibinfo {volume} {5}},\
  \bibinfo {pages} {2495} (\bibinfo {year} {2005})}\BibitemShut {NoStop}%
\bibitem [{\citenamefont {Malic}\ \emph {et~al.}(2018)\citenamefont {Malic},
  \citenamefont {Selig}, \citenamefont {Feierabend}, \citenamefont {Brem},
  \citenamefont {Christiansen}, \citenamefont {Wendler}, \citenamefont
  {Knorr},\ and\ \citenamefont {Bergh\"auser}}]{Malic:2018}%
  \BibitemOpen
  \bibfield  {author} {\bibinfo {author} {\bibfnamefont {E.}~\bibnamefont
  {Malic}}, \bibinfo {author} {\bibfnamefont {M.}~\bibnamefont {Selig}},
  \bibinfo {author} {\bibfnamefont {M.}~\bibnamefont {Feierabend}}, \bibinfo
  {author} {\bibfnamefont {S.}~\bibnamefont {Brem}}, \bibinfo {author}
  {\bibfnamefont {D.}~\bibnamefont {Christiansen}}, \bibinfo {author}
  {\bibfnamefont {F.}~\bibnamefont {Wendler}}, \bibinfo {author} {\bibfnamefont
  {A.}~\bibnamefont {Knorr}}, \ and\ \bibinfo {author} {\bibfnamefont
  {G.}~\bibnamefont {Bergh\"auser}},\ }\bibfield  {title} {\bibinfo {title}
  {Dark excitons in transition metal dichalcogenides},\ }\href {\doibase
  10.1103/PhysRevMaterials.2.014002} {\bibfield  {journal} {\bibinfo  {journal}
  {Phys. Rev. Materials}\ }\textbf {\bibinfo {volume} {2}},\ \bibinfo {pages}
  {014002} (\bibinfo {year} {2018})}\BibitemShut {NoStop}%
\bibitem [{\citenamefont {Stich}\ \emph {et~al.}(2014)\citenamefont {Stich},
  \citenamefont {Sp\"{a}th}, \citenamefont {Kraus}, \citenamefont {Sperlich},
  \citenamefont {Dyakonov},\ and\ \citenamefont {Hertel}}]{Stich:2014}%
  \BibitemOpen
  \bibfield  {author} {\bibinfo {author} {\bibfnamefont {D.}~\bibnamefont
  {Stich}}, \bibinfo {author} {\bibfnamefont {F.}~\bibnamefont {Sp\"{a}th}},
  \bibinfo {author} {\bibfnamefont {H.}~\bibnamefont {Kraus}}, \bibinfo
  {author} {\bibfnamefont {A.}~\bibnamefont {Sperlich}}, \bibinfo {author}
  {\bibfnamefont {V.}~\bibnamefont {Dyakonov}}, \ and\ \bibinfo {author}
  {\bibfnamefont {T.}~\bibnamefont {Hertel}},\ }\bibfield  {title} {\bibinfo
  {title} {Triplet-triplet exciton dynamics in single-walled carbon
  nanotubes},\ }\href {\doibase 10.1038/nphoton.2013.316} {\bibfield  {journal}
  {\bibinfo  {journal} {Nat. Photon.}\ }\textbf {\bibinfo {volume} {8}},\
  \bibinfo {pages} {139} (\bibinfo {year} {2014})}\BibitemShut {NoStop}%
\bibitem [{\citenamefont {Nishihara}\ \emph {et~al.}(2015)\citenamefont
  {Nishihara}, \citenamefont {Yamada}, \citenamefont {Okano},\ and\
  \citenamefont {Kanemitsu}}]{Nishihara:2015}%
  \BibitemOpen
  \bibfield  {author} {\bibinfo {author} {\bibfnamefont {T.}~\bibnamefont
  {Nishihara}}, \bibinfo {author} {\bibfnamefont {Y.}~\bibnamefont {Yamada}},
  \bibinfo {author} {\bibfnamefont {M.}~\bibnamefont {Okano}}, \ and\ \bibinfo
  {author} {\bibfnamefont {Y.}~\bibnamefont {Kanemitsu}},\ }\bibfield  {title}
  {\bibinfo {title} {Dynamics of the lowest-energy excitons in single-walled
  carbon nanotubes under resonant and nonresonant optical excitation},\ }\href
  {\doibase 10.1021/acs.jpcc.5b09485} {\bibfield  {journal} {\bibinfo
  {journal} {J. Phys. Chem. C}\ }\textbf {\bibinfo {volume} {119}},\ \bibinfo
  {pages} {28654} (\bibinfo {year} {2015})}\BibitemShut {NoStop}%
\bibitem [{\citenamefont {Zhang}\ \emph {et~al.}(2017)\citenamefont {Zhang},
  \citenamefont {Cao}, \citenamefont {Lu}, \citenamefont {Lin}, \citenamefont
  {Zhang}, \citenamefont {Wang}, \citenamefont {Li}, \citenamefont {Hone},
  \citenamefont {Robinson}, \citenamefont {Smirnov}, \citenamefont {Louie},\
  and\ \citenamefont {Heinz}}]{Zhang:2017}%
  \BibitemOpen
  \bibfield  {author} {\bibinfo {author} {\bibfnamefont {X.-X.}\ \bibnamefont
  {Zhang}}, \bibinfo {author} {\bibfnamefont {T.}~\bibnamefont {Cao}}, \bibinfo
  {author} {\bibfnamefont {Z.}~\bibnamefont {Lu}}, \bibinfo {author}
  {\bibfnamefont {Y.-C.}\ \bibnamefont {Lin}}, \bibinfo {author} {\bibfnamefont
  {F.}~\bibnamefont {Zhang}}, \bibinfo {author} {\bibfnamefont
  {Y.}~\bibnamefont {Wang}}, \bibinfo {author} {\bibfnamefont {Z.}~\bibnamefont
  {Li}}, \bibinfo {author} {\bibfnamefont {J.~C.}\ \bibnamefont {Hone}},
  \bibinfo {author} {\bibfnamefont {J.~A.}\ \bibnamefont {Robinson}}, \bibinfo
  {author} {\bibfnamefont {D.}~\bibnamefont {Smirnov}}, \bibinfo {author}
  {\bibfnamefont {S.~G.}\ \bibnamefont {Louie}}, \ and\ \bibinfo {author}
  {\bibfnamefont {T.~F.}\ \bibnamefont {Heinz}},\ }\bibfield  {title} {\bibinfo
  {title} {Magnetic brightening and control of dark excitons in monolayer
  {WS}e$_2$},\ }\href {https://doi.org/10.1038/nnano.2017.105} {\bibfield
  {journal} {\bibinfo  {journal} {Nat. Nanotech.}\ }\textbf {\bibinfo {volume}
  {12}},\ \bibinfo {pages} {883 } (\bibinfo {year} {2017})}\BibitemShut {NoStop}%
\bibitem [{\citenamefont {Zhou}\ \emph {et~al.}(2017)\citenamefont {Zhou},
  \citenamefont {Scuri}, \citenamefont {Wild}, \citenamefont {High},
  \citenamefont {Dibos}, \citenamefont {Jauregui}, \citenamefont {Shu},
  \citenamefont {De~Greve}, \citenamefont {Pistunova}, \citenamefont {Joe},
  \citenamefont {Taniguchi}, \citenamefont {Watanabe}, \citenamefont {Kim},
  \citenamefont {Lukin},\ and\ \citenamefont {Park}}]{Zhou:2017}%
  \BibitemOpen
  \bibfield  {author} {\bibinfo {author} {\bibfnamefont {Y.}~\bibnamefont
  {Zhou}}, \bibinfo {author} {\bibfnamefont {G.}~\bibnamefont {Scuri}},
  \bibinfo {author} {\bibfnamefont {D.~S.}\ \bibnamefont {Wild}}, \bibinfo
  {author} {\bibfnamefont {A.~A.}\ \bibnamefont {High}}, \bibinfo {author}
  {\bibfnamefont {A.}~\bibnamefont {Dibos}}, \bibinfo {author} {\bibfnamefont
  {L.~A.}\ \bibnamefont {Jauregui}}, \bibinfo {author} {\bibfnamefont
  {C.}~\bibnamefont {Shu}}, \bibinfo {author} {\bibfnamefont {K.}~\bibnamefont
  {De~Greve}}, \bibinfo {author} {\bibfnamefont {K.}~\bibnamefont {Pistunova}},
  \bibinfo {author} {\bibfnamefont {A.~Y.}\ \bibnamefont {Joe}}, \bibinfo
  {author} {\bibfnamefont {T.}~\bibnamefont {Taniguchi}}, \bibinfo {author}
  {\bibfnamefont {K.}~\bibnamefont {Watanabe}}, \bibinfo {author}
  {\bibfnamefont {P.}~\bibnamefont {Kim}}, \bibinfo {author} {\bibfnamefont
  {M.~D.}\ \bibnamefont {Lukin}}, \ and\ \bibinfo {author} {\bibfnamefont
  {H.}~\bibnamefont {Park}},\ }\bibfield  {title} {\bibinfo {title} {Probing
  dark excitons in atomically thin semiconductors via near-field coupling to
  surface plasmon polaritons},\ }\href {https://doi.org/10.1038/nnano.2017.106}
  {\bibfield  {journal} {\bibinfo  {journal} {Nat. Nanotech.}\ }\textbf
  {\bibinfo {volume} {12}},\ \bibinfo {pages} {856} (\bibinfo {year}
  {2017})}\BibitemShut {NoStop}%
\bibitem [{\citenamefont {Mortimer}\ and\ \citenamefont
  {Nicholas}(2007)}]{Mortimer:2007}%
  \BibitemOpen
  \bibfield  {author} {\bibinfo {author} {\bibfnamefont {I.~B.}\ \bibnamefont
  {Mortimer}}\ and\ \bibinfo {author} {\bibfnamefont {R.~J.}\ \bibnamefont
  {Nicholas}},\ }\bibfield  {title} {\bibinfo {title} {Role of bright and dark
  excitons in the temperature-dependent photoluminescence of carbon
  nanotubes},\ }\href {\doibase 10.1103/PhysRevLett.98.027404} {\bibfield
  {journal} {\bibinfo  {journal} {Phys. Rev. Lett.}\ }\textbf {\bibinfo
  {volume} {98}},\ \bibinfo {pages} {027404} (\bibinfo {year}
  {2007})}\BibitemShut {NoStop}%
\bibitem [{\citenamefont {Matsunaga}\ \emph {et~al.}(2009)\citenamefont
  {Matsunaga}, \citenamefont {Miyauchi}, \citenamefont {Matsuda},\ and\
  \citenamefont {Kanemitsu}}]{Matsunaga:2009}%
  \BibitemOpen
  \bibfield  {author} {\bibinfo {author} {\bibfnamefont {R.}~\bibnamefont
  {Matsunaga}}, \bibinfo {author} {\bibfnamefont {Y.}~\bibnamefont {Miyauchi}},
  \bibinfo {author} {\bibfnamefont {K.}~\bibnamefont {Matsuda}}, \ and\
  \bibinfo {author} {\bibfnamefont {Y.}~\bibnamefont {Kanemitsu}},\ }\bibfield
  {title} {\bibinfo {title} {Symmetry-induced nonequilibrium distributions of
  bright and dark exciton states in single carbon nanotubes},\ }\href {\doibase
  10.1103/PhysRevB.80.115436} {\bibfield  {journal} {\bibinfo  {journal} {Phys.
  Rev. B}\ }\textbf {\bibinfo {volume} {80}},\ \bibinfo {pages} {115436}
  (\bibinfo {year} {2009})}\BibitemShut {NoStop}%
\bibitem [{\citenamefont {Wang}\ \emph {et~al.}(2005)\citenamefont {Wang},
  \citenamefont {Dukovic}, \citenamefont {Brus},\ and\ \citenamefont
  {Heinz}}]{Wang:2005}%
  \BibitemOpen
  \bibfield  {author} {\bibinfo {author} {\bibfnamefont {F.}~\bibnamefont
  {Wang}}, \bibinfo {author} {\bibfnamefont {G.}~\bibnamefont {Dukovic}},
  \bibinfo {author} {\bibfnamefont {L.~E.}\ \bibnamefont {Brus}}, \ and\
  \bibinfo {author} {\bibfnamefont {T.~F.}\ \bibnamefont {Heinz}},\ }\bibfield
  {title} {\bibinfo {title} {The optical resonances in carbon nanotubes arise
  from excitons},\ }\href {\doibase 10.1126/science.1110265} {\bibfield
  {journal} {\bibinfo  {journal} {Science}\ }\textbf {\bibinfo {volume}
  {308}},\ \bibinfo {pages} {838} (\bibinfo {year} {2005})}\BibitemShut
  {NoStop}%
\bibitem [{\citenamefont {Maultzsch}\ \emph {et~al.}(2005)\citenamefont
  {Maultzsch}, \citenamefont {Pomraenke}, \citenamefont {Reich}, \citenamefont
  {Chang}, \citenamefont {Prezzi}, \citenamefont {Ruini}, \citenamefont
  {Molinari}, \citenamefont {Strano}, \citenamefont {Thomsen},\ and\
  \citenamefont {Lienau}}]{Maultzsch:2005}%
  \BibitemOpen
  \bibfield  {author} {\bibinfo {author} {\bibfnamefont {J.}~\bibnamefont
  {Maultzsch}}, \bibinfo {author} {\bibfnamefont {R.}~\bibnamefont
  {Pomraenke}}, \bibinfo {author} {\bibfnamefont {S.}~\bibnamefont {Reich}},
  \bibinfo {author} {\bibfnamefont {E.}~\bibnamefont {Chang}}, \bibinfo
  {author} {\bibfnamefont {D.}~\bibnamefont {Prezzi}}, \bibinfo {author}
  {\bibfnamefont {A.}~\bibnamefont {Ruini}}, \bibinfo {author} {\bibfnamefont
  {E.}~\bibnamefont {Molinari}}, \bibinfo {author} {\bibfnamefont {M.~S.}\
  \bibnamefont {Strano}}, \bibinfo {author} {\bibfnamefont {C.}~\bibnamefont
  {Thomsen}}, \ and\ \bibinfo {author} {\bibfnamefont {C.}~\bibnamefont
  {Lienau}},\ }\bibfield  {title} {\bibinfo {title} {Exciton binding energies
  in carbon nanotubes from two-photon photoluminescence},\ }\href {\doibase
  10.1103/PhysRevB.72.241402} {\bibfield  {journal} {\bibinfo  {journal} {Phys.
  Rev. B}\ }\textbf {\bibinfo {volume} {72}},\ \bibinfo {pages} {241402(R)}
  (\bibinfo {year} {2005})}\BibitemShut {NoStop}%
\bibitem [{\citenamefont {Luo}\ \emph {et~al.}(2015)\citenamefont {Luo},
  \citenamefont {Chatzakis}, \citenamefont {Patz},\ and\ \citenamefont
  {Wang}}]{Luo:2015}%
  \BibitemOpen
  \bibfield  {author} {\bibinfo {author} {\bibfnamefont {L.}~\bibnamefont
  {Luo}}, \bibinfo {author} {\bibfnamefont {I.}~\bibnamefont {Chatzakis}},
  \bibinfo {author} {\bibfnamefont {A.}~\bibnamefont {Patz}}, \ and\ \bibinfo
  {author} {\bibfnamefont {J.}~\bibnamefont {Wang}},\ }\bibfield  {title}
  {\bibinfo {title} {Ultrafast terahertz probes of interacting dark excitons in
  chirality-specific semiconducting single-walled carbon nanotubes},\ }\href
  {\doibase 10.1103/PhysRevLett.114.107402} {\bibfield  {journal} {\bibinfo
  {journal} {Phys. Rev. Lett.}\ }\textbf {\bibinfo {volume} {114}},\ \bibinfo
  {pages} {107402} (\bibinfo {year} {2015})}\BibitemShut {NoStop}%
\bibitem [{\citenamefont {Luo}\ \emph {et~al.}(2019)\citenamefont {Luo},
  \citenamefont {Liu}, \citenamefont {Yang}, \citenamefont {Vaswani},
  \citenamefont {Cheng}, \citenamefont {Park},\ and\ \citenamefont
  {Wang}}]{Luo:2019}%
  \BibitemOpen
  \bibfield  {author} {\bibinfo {author} {\bibfnamefont {L.}~\bibnamefont
  {Luo}}, \bibinfo {author} {\bibfnamefont {Z.}~\bibnamefont {Liu}}, \bibinfo
  {author} {\bibfnamefont {X.}~\bibnamefont {Yang}}, \bibinfo {author}
  {\bibfnamefont {C.}~\bibnamefont {Vaswani}}, \bibinfo {author} {\bibfnamefont
  {D.}~\bibnamefont {Cheng}}, \bibinfo {author} {\bibfnamefont {J.-M.}\
  \bibnamefont {Park}}, \ and\ \bibinfo {author} {\bibfnamefont
  {J.}~\bibnamefont {Wang}},\ }\bibfield  {title} {\bibinfo {title} {Anomalous
  variations of spectral linewidth in internal excitonic quantum transitions of
  ultrafast resonantly excited single-walled carbon nanotubes},\ }\href
  {\doibase 10.1103/PhysRevMaterials.3.026003} {\bibfield  {journal} {\bibinfo
  {journal} {Phys. Rev. Materials}\ }\textbf {\bibinfo {volume} {3}},\ \bibinfo
  {pages} {026003} (\bibinfo {year} {2019})}\BibitemShut {NoStop}%
\bibitem [{\citenamefont {Matsunaga}\ \emph {et~al.}(2008)\citenamefont
  {Matsunaga}, \citenamefont {Matsuda},\ and\ \citenamefont
  {Kanemitsu}}]{Matsunaga:2008}%
  \BibitemOpen
  \bibfield  {author} {\bibinfo {author} {\bibfnamefont {R.}~\bibnamefont
  {Matsunaga}}, \bibinfo {author} {\bibfnamefont {K.}~\bibnamefont {Matsuda}},
  \ and\ \bibinfo {author} {\bibfnamefont {Y.}~\bibnamefont {Kanemitsu}},\
  }\bibfield  {title} {\bibinfo {title} {Evidence for dark excitons in a single
  carbon nanotube due to the {A}haronov-{B}ohm effect},\ }\href {\doibase
  10.1103/PhysRevLett.101.147404} {\bibfield  {journal} {\bibinfo  {journal}
  {Phys. Rev. Lett.}\ }\textbf {\bibinfo {volume} {101}},\ \bibinfo {pages}
  {147404} (\bibinfo {year} {2008})}\BibitemShut {NoStop}%
\bibitem [{\citenamefont {Srivastava}\ \emph {et~al.}(2008)\citenamefont
  {Srivastava}, \citenamefont {Htoon}, \citenamefont {Klimov},\ and\
  \citenamefont {Kono}}]{Srivastava:2008}%
  \BibitemOpen
  \bibfield  {author} {\bibinfo {author} {\bibfnamefont {A.}~\bibnamefont
  {Srivastava}}, \bibinfo {author} {\bibfnamefont {H.}~\bibnamefont {Htoon}},
  \bibinfo {author} {\bibfnamefont {V.~I.}\ \bibnamefont {Klimov}}, \ and\
  \bibinfo {author} {\bibfnamefont {J.}~\bibnamefont {Kono}},\ }\bibfield
  {title} {\bibinfo {title} {Direct observation of dark excitons in individual
  carbon nanotubes: Inhomogeneity in the exchange splitting},\ }\href {\doibase
  10.1103/PhysRevLett.101.087402} {\bibfield  {journal} {\bibinfo  {journal}
  {Phys. Rev. Lett.}\ }\textbf {\bibinfo {volume} {101}},\ \bibinfo {pages}
  {087402} (\bibinfo {year} {2008})}\BibitemShut {NoStop}%
\bibitem [{\citenamefont {Uda}\ \emph {et~al.}(2016)\citenamefont {Uda},
  \citenamefont {Yoshida}, \citenamefont {Ishii},\ and\ \citenamefont
  {Kato}}]{Uda:2016}%
  \BibitemOpen
  \bibfield  {author} {\bibinfo {author} {\bibfnamefont {T.}~\bibnamefont
  {Uda}}, \bibinfo {author} {\bibfnamefont {M.}~\bibnamefont {Yoshida}},
  \bibinfo {author} {\bibfnamefont {A.}~\bibnamefont {Ishii}}, \ and\ \bibinfo
  {author} {\bibfnamefont {Y.~K.}\ \bibnamefont {Kato}},\ }\bibfield  {title}
  {\bibinfo {title} {Electric-field induced activation of dark excitonic states
  in carbon nanotubes},\ }\href {\doibase 10.1021/acs.nanolett.5b04595}
  {\bibfield  {journal} {\bibinfo  {journal} {Nano Lett.}\ }\textbf {\bibinfo
  {volume} {16}},\ \bibinfo {pages} {2278} (\bibinfo {year}
  {2016})}\BibitemShut {NoStop}%
\bibitem [{\citenamefont {Berciaud}\ \emph {et~al.}(2008)\citenamefont
  {Berciaud}, \citenamefont {Cognet},\ and\ \citenamefont
  {Lounis}}]{Berciaud:2008}%
  \BibitemOpen
  \bibfield  {author} {\bibinfo {author} {\bibfnamefont {S.}~\bibnamefont
  {Berciaud}}, \bibinfo {author} {\bibfnamefont {L.}~\bibnamefont {Cognet}}, \
  and\ \bibinfo {author} {\bibfnamefont {B.}~\bibnamefont {Lounis}},\
  }\bibfield  {title} {\bibinfo {title} {Luminescence decay and the absorption
  cross section of individual single-walled carbon nanotubes},\ }\href
  {\doibase 10.1103/PhysRevLett.101.077402} {\bibfield  {journal} {\bibinfo
  {journal} {Phys. Rev. Lett.}\ }\textbf {\bibinfo {volume} {101}},\ \bibinfo
  {pages} {077402} (\bibinfo {year} {2008})}\BibitemShut {NoStop}%
\bibitem [{\citenamefont {Gokus}\ \emph {et~al.}(2010)\citenamefont {Gokus},
  \citenamefont {Cognet}, \citenamefont {Duque}, \citenamefont {Pasquali},
  \citenamefont {Hartschuh},\ and\ \citenamefont {Lounis}}]{Gokus:2010}%
  \BibitemOpen
  \bibfield  {author} {\bibinfo {author} {\bibfnamefont {T.}~\bibnamefont
  {Gokus}}, \bibinfo {author} {\bibfnamefont {L.}~\bibnamefont {Cognet}},
  \bibinfo {author} {\bibfnamefont {J.~G.}\ \bibnamefont {Duque}}, \bibinfo
  {author} {\bibfnamefont {M.}~\bibnamefont {Pasquali}}, \bibinfo {author}
  {\bibfnamefont {A.}~\bibnamefont {Hartschuh}}, \ and\ \bibinfo {author}
  {\bibfnamefont {B.}~\bibnamefont {Lounis}},\ }\bibfield  {title} {\bibinfo
  {title} {Mono- and biexponential luminescence decays of individual
  single-walled carbon nanotubes},\ }\href {\doibase 10.1021/jp1049217}
  {\bibfield  {journal} {\bibinfo  {journal} {J. Phys. Chem. C}\ }\textbf
  {\bibinfo {volume} {114}},\ \bibinfo {pages} {14025} (\bibinfo {year}
  {2010})}\BibitemShut {NoStop}%
\bibitem [{\citenamefont {Hertel}\ \emph {et~al.}(2010)\citenamefont {Hertel},
  \citenamefont {Himmelein}, \citenamefont {Ackermann}, \citenamefont {Stich},\
  and\ \citenamefont {Crochet}}]{Hertel:2010}%
  \BibitemOpen
  \bibfield  {author} {\bibinfo {author} {\bibfnamefont {T.}~\bibnamefont
  {Hertel}}, \bibinfo {author} {\bibfnamefont {S.}~\bibnamefont {Himmelein}},
  \bibinfo {author} {\bibfnamefont {T.}~\bibnamefont {Ackermann}}, \bibinfo
  {author} {\bibfnamefont {D.}~\bibnamefont {Stich}}, \ and\ \bibinfo {author}
  {\bibfnamefont {J.}~\bibnamefont {Crochet}},\ }\bibfield  {title} {\bibinfo
  {title} {Diffusion limited photoluminescence quantum yields in 1-{D}
  semiconductors: Single-wall carbon nanotubes},\ }\href {\doibase
  10.1021/nn101612b} {\bibfield  {journal} {\bibinfo  {journal} {ACS Nano}\
  }\textbf {\bibinfo {volume} {4}},\ \bibinfo {pages} {7161} (\bibinfo {year}
  {2010})}\BibitemShut {NoStop}%
\bibitem [{\citenamefont {Crochet}\ \emph {et~al.}(2012)\citenamefont
  {Crochet}, \citenamefont {Duque}, \citenamefont {Werner}, \citenamefont
  {Lounis}, \citenamefont {Cognet},\ and\ \citenamefont
  {Doorn}}]{Crochet:2012}%
  \BibitemOpen
  \bibfield  {author} {\bibinfo {author} {\bibfnamefont {J.~J.}\ \bibnamefont
  {Crochet}}, \bibinfo {author} {\bibfnamefont {J.~G.}\ \bibnamefont {Duque}},
  \bibinfo {author} {\bibfnamefont {J.~H.}\ \bibnamefont {Werner}}, \bibinfo
  {author} {\bibfnamefont {B.}~\bibnamefont {Lounis}}, \bibinfo {author}
  {\bibfnamefont {L.}~\bibnamefont {Cognet}}, \ and\ \bibinfo {author}
  {\bibfnamefont {S.~K.}\ \bibnamefont {Doorn}},\ }\bibfield  {title} {\bibinfo
  {title} {Disorder limited exciton transport in colloidal single-wall carbon
  nanotubes},\ }\href {\doibase 10.1021/nl301739d} {\bibfield  {journal}
  {\bibinfo  {journal} {Nano Lett.}\ }\textbf {\bibinfo {volume} {12}},\
  \bibinfo {pages} {5091} (\bibinfo {year} {2012})}\BibitemShut {NoStop}%
\bibitem [{\citenamefont {Uda}\ \emph {et~al.}(2018)\citenamefont {Uda},
  \citenamefont {Tanaka},\ and\ \citenamefont {Kato}}]{Uda:2018a}%
  \BibitemOpen
  \bibfield  {author} {\bibinfo {author} {\bibfnamefont {T.}~\bibnamefont
  {Uda}}, \bibinfo {author} {\bibfnamefont {S.}~\bibnamefont {Tanaka}}, \ and\
  \bibinfo {author} {\bibfnamefont {Y.~K.}\ \bibnamefont {Kato}},\ }\bibfield
  {title} {\bibinfo {title} {Molecular screening effects on exciton-carrier
  interactions in suspended carbon nanotubes},\ }\href
  {https://doi.org/10.1063/1.5046433} {\bibfield  {journal} {\bibinfo
  {journal} {Appl. Phys. Lett.}\ }\textbf {\bibinfo {volume} {113}},\ \bibinfo
  {pages} {121105} (\bibinfo {year} {2018})}\BibitemShut {NoStop}%
\bibitem [{Sup()}]{SupplementalMaterials}%
  \BibitemOpen
  \href@noop {} {}\bibinfo {note} {See Supplemental Material for
  dependences of PL decay dynamics on excitation wavelength, detection
  wavelength, and excitation power, as well as state transtion from molecular
  desorbed state to adsorbed state.}\BibitemShut {Stop}%
\bibitem [{\citenamefont {Ishii}\ \emph {et~al.}(2015)\citenamefont {Ishii},
  \citenamefont {Yoshida},\ and\ \citenamefont {Kato}}]{Ishii:2015}%
  \BibitemOpen
  \bibfield  {author} {\bibinfo {author} {\bibfnamefont {A.}~\bibnamefont
  {Ishii}}, \bibinfo {author} {\bibfnamefont {M.}~\bibnamefont {Yoshida}}, \
  and\ \bibinfo {author} {\bibfnamefont {Y.~K.}\ \bibnamefont {Kato}},\
  }\bibfield  {title} {\bibinfo {title} {Exciton diffusion, end quenching, and
  exciton-exciton annihilation in individual air-suspended carbon nanotubes},\
  }\href {\doibase 10.1103/PhysRevB.91.125427} {\bibfield  {journal} {\bibinfo
  {journal} {Phys. Rev. B}\ }\textbf {\bibinfo {volume} {91}},\ \bibinfo
  {pages} {125427} (\bibinfo {year} {2015})}\BibitemShut {NoStop}%
\bibitem [{\citenamefont {Ishii}\ \emph {et~al.}(2018)\citenamefont {Ishii},
  \citenamefont {He}, \citenamefont {Hartmann}, \citenamefont {Machiya},
  \citenamefont {Htoon}, \citenamefont {Doorn},\ and\ \citenamefont
  {Kato}}]{Ishii:2018}%
  \BibitemOpen
  \bibfield  {author} {\bibinfo {author} {\bibfnamefont {A.}~\bibnamefont
  {Ishii}}, \bibinfo {author} {\bibfnamefont {X.}~\bibnamefont {He}}, \bibinfo
  {author} {\bibfnamefont {N.~F.}\ \bibnamefont {Hartmann}}, \bibinfo {author}
  {\bibfnamefont {H.}~\bibnamefont {Machiya}}, \bibinfo {author} {\bibfnamefont
  {H.}~\bibnamefont {Htoon}}, \bibinfo {author} {\bibfnamefont {S.~K.}\
  \bibnamefont {Doorn}}, \ and\ \bibinfo {author} {\bibfnamefont {Y.~K.}\
  \bibnamefont {Kato}},\ }\bibfield  {title} {\bibinfo {title} {Enhanced
  single-photon emission from carbon-nanotube dopant states coupled to silicon
  microcavities},\ }\href {\doibase 10.1021/acs.nanolett.8b01170} {\bibfield
  {journal} {\bibinfo  {journal} {Nano Lett.}\ }\textbf {\bibinfo {volume}
  {18}},\ \bibinfo {pages} {3873} (\bibinfo {year} {2018})}\BibitemShut
  {NoStop}%
\bibitem [{\citenamefont {Anderson}\ \emph {et~al.}(2013)\citenamefont
  {Anderson}, \citenamefont {Xiao},\ and\ \citenamefont
  {Fraser}}]{Anderson:2013}%
  \BibitemOpen
  \bibfield  {author} {\bibinfo {author} {\bibfnamefont {M.~D.}\ \bibnamefont
  {Anderson}}, \bibinfo {author} {\bibfnamefont {Y.-F.}\ \bibnamefont {Xiao}},
  \ and\ \bibinfo {author} {\bibfnamefont {J.~M.}\ \bibnamefont {Fraser}},\
  }\bibfield  {title} {\bibinfo {title} {First-passage theory of exciton
  population loss in single-walled carbon nanotubes reveals micron-scale
  intrinsic diffusion lengths},\ }\href {\doibase 10.1103/PhysRevB.88.045420}
  {\bibfield  {journal} {\bibinfo  {journal} {Phys. Rev. B}\ }\textbf {\bibinfo
  {volume} {88}},\ \bibinfo {pages} {045420} (\bibinfo {year}
  {2013})}\BibitemShut {NoStop}%
\bibitem [{\citenamefont {Xiao}\ \emph {et~al.}(2010)\citenamefont {Xiao},
  \citenamefont {Nhan}, \citenamefont {Wilson},\ and\ \citenamefont
  {Fraser}}]{Xiao:2010}%
  \BibitemOpen
  \bibfield  {author} {\bibinfo {author} {\bibfnamefont {Y.-F.}\ \bibnamefont
  {Xiao}}, \bibinfo {author} {\bibfnamefont {T.~Q.}\ \bibnamefont {Nhan}},
  \bibinfo {author} {\bibfnamefont {M.~W.~B.}\ \bibnamefont {Wilson}}, \ and\
  \bibinfo {author} {\bibfnamefont {J.~M.}\ \bibnamefont {Fraser}},\ }\bibfield
   {title} {\bibinfo {title} {Saturation of the photoluminescence at
  few-exciton levels in a single-walled carbon nanotube under ultrafast
  excitation},\ }\href {\doibase 10.1103/PhysRevLett.104.017401} {\bibfield
  {journal} {\bibinfo  {journal} {Phys. Rev. Lett.}\ }\textbf {\bibinfo
  {volume} {104}},\ \bibinfo {pages} {017401} (\bibinfo {year}
  {2010})}\BibitemShut {NoStop}%
\bibitem [{\citenamefont {Moritsubo}\ \emph {et~al.}(2010)\citenamefont
  {Moritsubo}, \citenamefont {Murai}, \citenamefont {Shimada}, \citenamefont
  {Murakami}, \citenamefont {Chiashi}, \citenamefont {Maruyama},\ and\
  \citenamefont {Kato}}]{Moritsubo:2010}%
  \BibitemOpen
  \bibfield  {author} {\bibinfo {author} {\bibfnamefont {S.}~\bibnamefont
  {Moritsubo}}, \bibinfo {author} {\bibfnamefont {T.}~\bibnamefont {Murai}},
  \bibinfo {author} {\bibfnamefont {T.}~\bibnamefont {Shimada}}, \bibinfo
  {author} {\bibfnamefont {Y.}~\bibnamefont {Murakami}}, \bibinfo {author}
  {\bibfnamefont {S.}~\bibnamefont {Chiashi}}, \bibinfo {author} {\bibfnamefont
  {S.}~\bibnamefont {Maruyama}}, \ and\ \bibinfo {author} {\bibfnamefont
  {Y.~K.}\ \bibnamefont {Kato}},\ }\bibfield  {title} {\bibinfo {title}
  {Exciton diffusion in air-suspended single-walled carbon nanotubes},\ }\href
  {\doibase 10.1103/PhysRevLett.104.247402} {\bibfield  {journal} {\bibinfo
  {journal} {Phys. Rev. Lett.}\ }\textbf {\bibinfo {volume} {104}},\ \bibinfo
  {pages} {247402} (\bibinfo {year} {2010})}\BibitemShut {NoStop}%
\bibitem [{\citenamefont {Srivastava}\ and\ \citenamefont
  {Kono}(2009)}]{Srivastava:2009}%
  \BibitemOpen
  \bibfield  {author} {\bibinfo {author} {\bibfnamefont {A.}~\bibnamefont
  {Srivastava}}\ and\ \bibinfo {author} {\bibfnamefont {J.}~\bibnamefont
  {Kono}},\ }\bibfield  {title} {\bibinfo {title} {Diffusion-limited
  exciton-exciton annihilation in single-walled carbon nanotubes: A
  time-dependent analysis},\ }\href {\doibase 10.1103/PhysRevB.79.205407}
  {\bibfield  {journal} {\bibinfo  {journal} {Phys. Rev. B}\ }\textbf {\bibinfo
  {volume} {79}},\ \bibinfo {pages} {205407} (\bibinfo {year}
  {2009})}\BibitemShut {NoStop}%
\bibitem [{\citenamefont {Capaz}\ \emph {et~al.}(2006)\citenamefont {Capaz},
  \citenamefont {Spataru}, \citenamefont {Ismail-Beigi},\ and\ \citenamefont
  {Louie}}]{Capaz:2006}%
  \BibitemOpen
  \bibfield  {author} {\bibinfo {author} {\bibfnamefont {R.~B.}\ \bibnamefont
  {Capaz}}, \bibinfo {author} {\bibfnamefont {C.~D.}\ \bibnamefont {Spataru}},
  \bibinfo {author} {\bibfnamefont {S.}~\bibnamefont {Ismail-Beigi}}, \ and\
  \bibinfo {author} {\bibfnamefont {S.~G.}\ \bibnamefont {Louie}},\ }\bibfield
  {title} {\bibinfo {title} {Diameter and chirality dependence of exciton
  properties in carbon nanotubes},\ }\href {\doibase
  10.1103/PhysRevB.74.121401} {\bibfield  {journal} {\bibinfo  {journal} {Phys.
  Rev. B}\ }\textbf {\bibinfo {volume} {74}},\ \bibinfo {pages} {121401(R)}
  (\bibinfo {year} {2006})}\BibitemShut {NoStop}%
\bibitem [{\citenamefont {Lefebvre}\ and\ \citenamefont
  {Finnie}(2008)}]{Lefebvre:2008}%
  \BibitemOpen
  \bibfield  {author} {\bibinfo {author} {\bibfnamefont {J.}~\bibnamefont
  {Lefebvre}}\ and\ \bibinfo {author} {\bibfnamefont {P.}~\bibnamefont
  {Finnie}},\ }\bibfield  {title} {\bibinfo {title} {Excited excitonic states
  in single-walled carbon nanotubes},\ }\href {\doibase 10.1021/nl080518h}
  {\bibfield  {journal} {\bibinfo  {journal} {Nano Lett.}\ }\textbf {\bibinfo
  {volume} {8}},\ \bibinfo {pages} {1890} (\bibinfo {year} {2008})}\BibitemShut
  {NoStop}%
\bibitem [{\citenamefont {Ma}\ \emph {et~al.}(2015{\natexlab{a}})\citenamefont
  {Ma}, \citenamefont {Roslyak}, \citenamefont {Duque}, \citenamefont {Pang},
  \citenamefont {Doorn}, \citenamefont {Piryatinski}, \citenamefont {Dunlap},\
  and\ \citenamefont {Htoon}}]{Ma:2015prl}%
  \BibitemOpen
  \bibfield  {author} {\bibinfo {author} {\bibfnamefont {X.}~\bibnamefont
  {Ma}}, \bibinfo {author} {\bibfnamefont {O.}~\bibnamefont {Roslyak}},
  \bibinfo {author} {\bibfnamefont {J.~G.}\ \bibnamefont {Duque}}, \bibinfo
  {author} {\bibfnamefont {X.}~\bibnamefont {Pang}}, \bibinfo {author}
  {\bibfnamefont {S.~K.}\ \bibnamefont {Doorn}}, \bibinfo {author}
  {\bibfnamefont {A.}~\bibnamefont {Piryatinski}}, \bibinfo {author}
  {\bibfnamefont {D.~H.}\ \bibnamefont {Dunlap}}, \ and\ \bibinfo {author}
  {\bibfnamefont {H.}~\bibnamefont {Htoon}},\ }\bibfield  {title} {\bibinfo
  {title} {Influences of exciton diffusion and exciton-exciton annihilation on
  photon emission statistics of carbon nanotubes},\ }\href {\doibase
  10.1103/PhysRevLett.115.017401} {\bibfield  {journal} {\bibinfo  {journal}
  {Phys. Rev. Lett.}\ }\textbf {\bibinfo {volume} {115}},\ \bibinfo {pages}
  {017401} (\bibinfo {year} {2015}{\natexlab{a}})}\BibitemShut {NoStop}%
\bibitem [{\citenamefont {Ma}\ \emph {et~al.}(2015{\natexlab{b}})\citenamefont
  {Ma}, \citenamefont {Hartmann}, \citenamefont {Baldwin}, \citenamefont
  {Doorn},\ and\ \citenamefont {Htoon}}]{Ma:2015NatNano}%
  \BibitemOpen
  \bibfield  {author} {\bibinfo {author} {\bibfnamefont {X.}~\bibnamefont
  {Ma}}, \bibinfo {author} {\bibfnamefont {N.~F.}\ \bibnamefont {Hartmann}},
  \bibinfo {author} {\bibfnamefont {J.~K.~S.}\ \bibnamefont {Baldwin}},
  \bibinfo {author} {\bibfnamefont {S.~K.}\ \bibnamefont {Doorn}}, \ and\
  \bibinfo {author} {\bibfnamefont {H.}~\bibnamefont {Htoon}},\ }\bibfield
  {title} {\bibinfo {title} {Room-temperature single-photon generation from
  solitary dopants of carbon nanotubes},\ }\href {\doibase
  10.1038/nnano.2015.136} {\bibfield  {journal} {\bibinfo  {journal} {Nat.
  Nanotech.}\ }\textbf {\bibinfo {volume} {10}},\ \bibinfo {pages} {671}
  (\bibinfo {year} {2015}{\natexlab{b}})}\BibitemShut {NoStop}%
\bibitem [{\citenamefont {Furchi}\ \emph {et~al.}(2014)\citenamefont {Furchi},
  \citenamefont {Pospischil}, \citenamefont {Libisch}, \citenamefont
  {Burgdorfer},\ and\ \citenamefont {Mueller}}]{Furchi:2014}%
  \BibitemOpen
  \bibfield  {author} {\bibinfo {author} {\bibfnamefont {M.~M.}\ \bibnamefont
  {Furchi}}, \bibinfo {author} {\bibfnamefont {A.}~\bibnamefont {Pospischil}},
  \bibinfo {author} {\bibfnamefont {F.}~\bibnamefont {Libisch}}, \bibinfo
  {author} {\bibfnamefont {J.}~\bibnamefont {Burgdorfer}}, \ and\ \bibinfo
  {author} {\bibfnamefont {T.}~\bibnamefont {Mueller}},\ }\bibfield  {title}
  {\bibinfo {title} {Photovoltaic effect in an electrically tunable van der
  {W}aals heterojunction},\ }\href {\doibase 10.1021/nl501962c} {\bibfield
  {journal} {\bibinfo  {journal} {Nano Lett.}\ }\textbf {\bibinfo {volume}
  {14}},\ \bibinfo {pages} {4785} (\bibinfo {year} {2014})}\BibitemShut
  {NoStop}%
\bibitem [{\citenamefont {Memaran}\ \emph {et~al.}(2015)\citenamefont
  {Memaran}, \citenamefont {Pradhan}, \citenamefont {Lu}, \citenamefont
  {Rhodes}, \citenamefont {Ludwig}, \citenamefont {Zhou}, \citenamefont
  {Ogunsolu}, \citenamefont {Ajayan}, \citenamefont {Smirnov}, \citenamefont
  {Fernandez-Dominguez}, \citenamefont {Garcia-Vidal},\ and\ \citenamefont
  {Balicas}}]{Memaran:2015}%
  \BibitemOpen
  \bibfield  {author} {\bibinfo {author} {\bibfnamefont {S.}~\bibnamefont
  {Memaran}}, \bibinfo {author} {\bibfnamefont {N.~R.}\ \bibnamefont
  {Pradhan}}, \bibinfo {author} {\bibfnamefont {Z.}~\bibnamefont {Lu}},
  \bibinfo {author} {\bibfnamefont {D.}~\bibnamefont {Rhodes}}, \bibinfo
  {author} {\bibfnamefont {J.}~\bibnamefont {Ludwig}}, \bibinfo {author}
  {\bibfnamefont {Q.}~\bibnamefont {Zhou}}, \bibinfo {author} {\bibfnamefont
  {O.}~\bibnamefont {Ogunsolu}}, \bibinfo {author} {\bibfnamefont {P.~M.}\
  \bibnamefont {Ajayan}}, \bibinfo {author} {\bibfnamefont {D.}~\bibnamefont
  {Smirnov}}, \bibinfo {author} {\bibfnamefont {A.~I.}\ \bibnamefont
  {Fernandez-Dominguez}}, \bibinfo {author} {\bibfnamefont {F.~J.}\
  \bibnamefont {Garcia-Vidal}}, \ and\ \bibinfo {author} {\bibfnamefont
  {L.}~\bibnamefont {Balicas}},\ }\bibfield  {title} {\bibinfo {title}
  {Pronounced photovoltaic response from multilayered transition-metal
  dichalcogenides {PN}-junctions},\ }\href {\doibase
  10.1021/acs.nanolett.5b03265} {\bibfield  {journal} {\bibinfo  {journal}
  {Nano Lett.}\ }\textbf {\bibinfo {volume} {15}},\ \bibinfo {pages} {7532}
  (\bibinfo {year} {2015})}\BibitemShut {NoStop}%
\end{thebibliography}
\end{document}